# Highly efficient mid-infrared on-chip electrical generation of graphene plasmons by inelastic electron tunnelling excitation

Kelvin J. A. Ooi, [1,*] H. S. Chu, [2,*] C. Y. Hsieh, [1,3] Dawn T. H. Tan, [1] and L. K. Ang[1]

[1]Engineering Product Development, Singapore University of Technology and Design, East Coast Campus, 8 Somapah Road, Singapore 487372

[2]Electronics & Photonics Department, A*STAR Institute of High Performance Computing, 1 Fusionopolis Way, #16-16 Connexis, Singapore 138632

[3]Department of Chemistry, Massachusetts Institute of Technology, Cambridge, Massachusetts 02139, USA

*Corresponding author: kelvin_ooi@sutd.edu.sg; chuhs@ihpc.a-star.edu.sg

**Abstract:** Inelastic electron tunnelling provides a low energy pathway for the excitation of surface plasmons and light emission. We theoretically investigated tunnel-junctions based on metals and graphene. We show that graphene is potentially a highly-efficient material for tunnelling excitation of plasmons, thanks to its narrow plasmon linewidths, strong emission and large tunability in the mid-infrared wavelength regime. Compared to gold and silver, the enhancement can be up to 10 times for similar wavelengths, and up to 5 orders at their respective plasmon operating wavelengths. Tunnelling-excitation of graphene plasmons promises an efficient technology for on-chip electrical generation and manipulation of plasmons for graphene-based optoelectronics and nanophotonic integrated circuits.

## I. Introduction

In 1976, Lambe and McCarthy discovered light emission from metal-insulator-metal (MIM) tunnel junctions [1]. Following the discovery, light emitting tunnel-junctions (LETJs) had drawn research interests in both theory and experiment to understand the emission process [2–5]. The general consensus in those literature is that the light emission is plasmon-mediated. Recently, there are renewed interests in the tunnel-junction as surface plasmon sources [6–11]. One of the advantage of having surface plasmon tunnel junctions (SPTJs) is the in-situ generation of surface plasmons, avoiding the need for free-space optical coupling which are often bulky and inefficient. Another advantage for SPTJs is the low-energy excitation compared to other higher-energy excitation pathways like electron-bombardment or aloof-scattering [12–14].

Existing LETJs and SPTJs used metals as the tunnel-junction electrodes, in particular, aluminium, silver and gold [2–11]. In this paper, we theoretically study graphene as a material for the tunnelling-excitation of surface plasmons. Most of the current researches on graphene tunnel junctions show that inelastic electron tunnelling occurs mainly through phonon channels [15–17]. However, inelastic electron tunnelling through plasmon channels has also been experimentally demonstrated before [15]. We will show that thanks to graphene's narrow plasmon linewidths, and SPPs operating in the mid-infrared wavelength regime, the inelastic electron tunnelling excitation of graphene plasmons will be much more efficient compared to most metals.

The prevailing theory for the inelastic electron tunneling plasmon (IETP) excitation mechanism is that at low bias voltages, it occurs through a three-step process [5–11]:

   i)     electrons tunnel inelastically from one electrode to another in an MIM junction,

ii) gap plasmons are excited through coupling from the energy-loss ℏω,
iii) gap plasmons are coupled out to surface plasmon polaritons (SPPs) and then into radiation.

Hence, we investigate IETP excitation using a two-part approach: in the first part, we studied the frequency-dependent gap plasmon formation from IETP excitation using available formulations [2]; then, in the second part, we perform the finite-difference time-domain (FDTD) simulations by means of the Lumerical commercial software [18], to find the coupling efficiency of gap plasmons into SPPs. The final result is obtained by superimposing the optical power spectrum from both parts. Our approach could fairly reproduce some of the spectral features from previous IETP excitation experiments [4,6,10].

## II. Theory of gap plasmon excitation from inelastic electron tunnelling
### A. Gap plasmons in metal-insulator-metal (MIM) tunnel junctions

To find the gap plasmon power spectrum, firstly we need to find the tunnelling currents and induced gap plasmon electric-fields. The formulation for an MIM tunnel junction depicted in Fig. 1(a) can be found in ref. [2], but will be briefly reproduced here. The tunnelling current is written as

$$J = \left(\frac{ie\hbar}{2m_e^*}\right)\left(\psi_R^* \nabla \psi_L - \psi_L \nabla \psi_R^*\right) \tag{1}$$

where ψ is the electronic wave-function on the left (L) and right (R) electrodes and are described by

$$\psi_L = \frac{1}{\sqrt{2Al}} \chi_L \exp(-iE_L t/\hbar) \tag{2a}$$

$$\psi_R = \frac{1}{\sqrt{2Al}} \chi_R \exp(-i[qx + E_R t/\hbar]) \tag{2b}$$

where A and *l* are the normalization of the area and length of the electrodes respectively, E is the energy of the electrodes, q is the plasmon vector, and χ is the eigenfunction given by [19]

$$\chi_L = \begin{cases} e^{ik_L z} + R_L e^{-ik_L z} &, \quad z < 0 \\ C_L e^{-K_L z} + D_L e^{K_L z} &, \quad 0 < z < d \\ T_L e^{ik_L(z-d)} &, \quad d < z \end{cases} \tag{2c}$$

$$\chi_R = \begin{cases} T_R e^{-ik_R z} &, \quad z < 0 \\ C_R e^{-K_R(d-z)} + D_R e^{K_R(d-z)} &, \quad 0 < z < d \\ e^{-ik_R(z-d)} + R_R e^{-ik_R(z-d)} &, \quad d < z \end{cases} \tag{2d}$$

where R, C, D and T are the coefficients found from the boundary conditions such that χ and dχ/dz are continuous at z=0 and z=d. The eigenfunctions carry the momentum terms

$$k_L = \left(2m_e^* E_L/\hbar^2\right)^{1/2} \tag{3a}$$

$$K_L = \left(2m_e^*(U_0 - E_L)/\hbar^2\right)^{1/2} \tag{3b}$$

$$k_R = \left(2m_e^* E_R/\hbar^2 - q^2\right)^{1/2} \tag{3c}$$

$$K_R = \left(2m_e^*(U_0 - E_R)/\hbar^2 + q^2\right)^{1/2} \tag{3d}$$

where $U_0$ is the barrier height of the emitting left electrode, and $m_e^*$ is the effective electron mass.

Figure 2(a) depicts the inelastic electron tunnelling process used to generate plasmons. In the equilibrium state, $E_L$ and $E_R$ align at the same Fermi level. On the application of a bias voltage across the MIM junction, the energy levels shift and the difference in Fermi level drives the tunnelling currents and then excite the gap plasmon, with energy $E_L–E_R=\hbar\omega$. The cut-off frequency is determined by the applied bias-voltage, $\hbar\omega=eV_z$ [1].

Then, the induced electric-fields, arising from the transition charges, could be simply found from Gauss law

$$\nabla \cdot [\varepsilon_{(z,\omega)} \nabla \phi] = 4\pi \rho_z \tag{4}$$

where $\varepsilon$ is the permittivity of the material which is dependent on frequency $\omega$ and location z, $\rho = -e\psi_L \psi_R^*$ is the charge density distribution, and $F = -\nabla\phi$ is the electric field. Finally, the gap plasmon power could be obtained from the equation

$$P_{gap-plasmon} = -2\,\mathrm{Re}\int F^* \cdot J\,dz \tag{5}$$

Next, the surface plasmon spectrum could be obtained by defining a frequency-dependent q, which is obtained from the surface plasmon dispersion curve for MIM structures given by

$$(\varepsilon_L + \varepsilon_0)(\varepsilon_R + \varepsilon_0) - (\varepsilon_L - \varepsilon_0)(\varepsilon_R - \varepsilon_0)\exp(-2qd) = 0 \tag{6}$$

where $\varepsilon_L$, $\varepsilon_R$ and $\varepsilon_0$ are the permittivity for the left and right electrodes, and the gap respectively.

### B. Gap plasmons in metal-insulator-graphene (MIG) tunnel junctions

The development of the theoretical formulation of the graphene IETP system is similar to the ones described from the previous section, but with a few modifications. Firstly, there's a tunnelling electron mass anisotropy in graphene [20], where the out-of-plane mass $m_\perp = m_e$ is normal while the in-plane mass is relativistic, given by $m_\parallel = E_F/v_F^2$ [21], where $E_F$ is the Fermi level of graphene while $v_F = 10^6$ m/s is the Fermi velocity. This results in modification of the wave-function momentum-terms at the right interface

$$k_R = \left(\frac{2m_\perp E_R}{\hbar^2} - \frac{m_\perp}{m_\parallel}q^2\right)^{1/2} \tag{7a}$$

$$K_R = \left(\frac{2m_\perp(U_0 - E_R)}{\hbar^2} + \frac{m_\perp}{m_\parallel}q^2\right)^{1/2} \tag{7b}$$

Next, as shown in Fig. 1(b), the graphene is modelled as a 2D material, and, therefore, it has an associated 2D charge density. This will result in the boundary conditions for the electric-fields, $\varepsilon\nabla\phi_z$ being not continuous at z=d.

We start by writing the solutions for the electric-potential boundary conditions

$$\phi(z) = \begin{cases} a_L e^{qz} + g_L(z) & , \quad z < 0 \\ a_+ e^{qz} + a_- e^{-qz} + g_0(z) & , \quad 0 < z < d \\ a_R e^{-q(z-d)} + g_R(z) & , \quad d < z \end{cases} \qquad (8)$$

where $g_L$, $g_0$ and $g_R$ are the inhomogeneous parts of the solutions [2]. Hence we can write the algebraic equations

$$a_L + g_L(0) = a_+ + a_- + g_0(0) \qquad (9a)$$

$$a_+ e^{qd} + a_- e^{-qd} + g_0(d) = a_R + g_R(d) \qquad (9b)$$

Since $\varepsilon\nabla\phi_z$ is not continuous at z=d, the charge-density term $i\omega\rho = \nabla \cdot J_x$ arises, and

$$\begin{aligned} \nabla \cdot J_x &= \sigma \nabla^2 \phi_{(z=d)} \\ &= -\sigma q^2 \phi_{(z=d)} \end{aligned} \qquad (10)$$

where $\sigma$ is the 2D conductivity of graphene. Thus, we write the algebraic equations for the electric-fields:

$$\varepsilon_L [a_L q + g'_L(0)] = \varepsilon_0 [a_+ q - a_- q + g'_0(0)] \qquad (11a)$$

$$\varepsilon_0 [a_+ e^{qd} q - a_- e^{-qd} q + g'_0(d)] = \varepsilon_R [-a_R q + g'_R(d)] - \frac{iq^2\sigma}{\omega}[a_+ e^{qd} + a_- e^{-qd} + g_0(d)] \qquad (11b)$$

Combining and rearranging all the 4 equations from (9a), (9b), (11a) and (11b), we find the coefficients

$$a_L = a_+ + a_- + g_0(0) - g_L(0) \qquad (12a)$$

$$a_R = a_+ e^{qb} + a_- e^{-qb} + g_0(b) - g_R(b) \qquad (12b)$$

$$a_+ = -e^{-qd}\left[e^{-qd} S_1(\varepsilon_R - \varepsilon_0 + iq\sigma/\omega) - S_2(\varepsilon_L + \varepsilon_0)\right]/\eta \qquad (12c)$$

$$a_- = e^{-qd}\left[e^{qd} S_1(\varepsilon_0 + \varepsilon_R + iq\sigma/\omega) - S_2(\varepsilon_L - \varepsilon_0)\right]/\eta \qquad (12d)$$

where

$$\eta = (\varepsilon_L + \varepsilon_0)(\varepsilon_R + \varepsilon_0 + iq\sigma/\omega) - (\varepsilon_L - \varepsilon_0)(\varepsilon_R - \varepsilon_0 + iq\sigma/\omega)e^{-2qd} \qquad (13a)$$

$$S_1 = \varepsilon_L[g_L(0) - g_0(0)] + [\varepsilon_0 g'_0(0) - \varepsilon_L g'_L(0)]/q \qquad (13b)$$

$$S_2 = \varepsilon_R[g_R(d) - g_0(d)] + [\varepsilon_R g'_R(d) - \varepsilon_0 g'_0(d)]/q - iq\sigma g_0(b)/\omega \qquad (13c)$$

The 2D charge density allows the formation of the gap plasmon between graphene and the metal tip. Here, η also represents the pole of the p-polarized reflection coefficient of the MIG junction. Consequently, we can find the dispersion for MIG tunnel junctions by letting η→0.

### III. Verification models for experimental SPP and emission spectra

Following our developed theoretical approach as stated in section I and II, we will try to reproduce the IETP excitation spectra from past experiments. The gap plasmon spectrum is numerically obtained from formulations in section II, while the gap plasmon to SPP coupling spectrum is obtained from Lumerical 2D FDTD simulations. In the FDTD simulation, the gap plasmon is modelled as an electric dipole source, with power normalized at 1W at all wavelengths, connecting the emitter electrode and the metal substrate as shown in Fig. 2(b). The gap plasmons would then couple to the SPP modes and propagate along the metal substrate. The SPP is allowed to propagate a few micrometers along the substrate, and then the near-field spectrum of the SPP is recorded. Superimposing both spectra would result in the SPP or emission intensity spectrum.

In our first study, we simulated the intensity spectrum of an Al-$Al_2O_3$-Ag tunnel junction for different bias-voltages as described in ref. [4]. The complex permittivity for Al is taken from Palik's handbook [22], while that for Ag is obtained from Johnson and Christy's paper [23]. The permittivity of $Al_2O_3$ is approximately 3, and has an effective tunnelling mass of $0.2 \times m_e$ [24]. The results of both simulated and experimental spectra are compared in Fig. 3. It is observed that the simulated results could reproduce the general shape of the spectra as well as the cutoff wavelengths of the emission. However, there are some discrepancies in the peak wavelengths, where the simulated peak wavelength for 2.7V is slightly red-shifted (575nm) compared to the experimental peak (540nm). The discrepancy gets larger at higher bias-voltages. The blue-shift of the experimental spectra is due to the spectra being recorded in the far-field, in contrast to our simulation spectra being recorded in the near-field [25]. Moreover, the discrepancies could also be due to the imperfection of the fabricated structures such as surface roughness, and the value of the complex permittivity used in modelling. Nevertheless, the IETP excitation theory is adequate for comparative analysis of excitation efficiencies between different plasmonic materials.

In our second study, we simulate an Au-air-Au IETP system biased at 2V, as described in ref. [6]. In our FDTD simulation structure, the Au tip radius is 50nm. The permittivity of Au is obtained from Johnson and Christy [23]. The simulated spectrum in Fig. 4(a) shows that the spectrum shape and cutoff wavelength have good agreement with the experimental results from ref. [6]. Similar to our first study, the simulated Au-air-Au IETP system has a red-shifted peak wavelength (780nm) compared to the experimental spectrum (720nm).

Our third study consists of the same Au-air-Au IETP system but under a bias of 2.5V. With a higher bias-voltage, the peak wavelength (750nm) of the simulated spectrum in Fig. 4(b) is shorter compared to the second study. Compared to the experimental results from ref. [10] (700nm), the obtained result shows a similar red-shifted peak.

### IV. Comparison of IETP performance for metals and graphene

We numerically calculated the IETP gap plasmon power for tunnel-junctions between an aluminium tip and samples consisting of metals and graphene, for a tip-bias of 2V. For metals, we selected gold and silver, which have good plasmonic properties in the visible regime. While for graphene, we studied the case for four doped-graphene Fermi-levels ranging from 0.1–0.4eV. The conductivity and material parameters for graphene are taken from ref. [25]. In all cases, a 3nm air gap between the tip and sample (d=3nm) is chosen, and graphene is assumed to be free-standing for simplicity. From the calculated results plotted in Fig. 5, we predicted that IETP gap plasmon power for graphene is very much larger than that for gold and silver by up to 10x in the long wavelength regime. The IETP gap plasmon power rises quickly from the cut-off wavelength (defined by the lower between the tip-bias cutoff frequency and the gap plasmon resonance frequency), and saturates at the long wavelength regime.

However, not all wavelengths support the formation and sustenance of SPPs. Full assessment of the SPP generation efficiency requires knowledge of the coupling efficiency of gap plasmons into SPPs. Here, we perform Lumerical 2D FDTD simulations of the IETP systems as depicted in Fig. 1(a) and (b) as a direct way to obtain the coupling efficiencies. In our simulations, we standardized the structure to be an infinitely-long and straight aluminium flat tip of 20nm width, placed 3nm away in air from the metal and graphene substrate, as depicted in Fig. 6(a). A dipole excitation source is placed in the middle of the air gap to simulate the gap plasmon, and the out-coupled SPP power in the x-direction is recorded. It is found that the coupling efficiency (normalized to propagation losses) of the gap plasmons to SPPs on gold, silver and graphene substrates are between 0.5–2%, as shown in Fig. 6(b)–(d). However, the critical difference lies in their respective SPP operating wavelengths, i.e. gold and silver in the visible regime, and graphene in the mid-infrared regime. Importantly, if we also take the gap plasmon generation efficiency into account (by superimposing Fig. 6(b)–(d) with Fig. 5), we find that the total generation efficiency of graphene SPPs greatly surpasses that of gold and silver by 5–6 orders, as shown in Fig. 6(e)–(g).

There are at least two reasons why the IETP efficiency of graphene plasmons is greater than that of metal plasmons. The first reason, as stated before, is the different operating wavelength regimes. It is easier for electrons to lose energy inelastically through lower plasmon momentum channels compared to higher ones.

The second reason could be attributed to the narrow linewidths of the generated graphene plasmons. To examine the plasmon linewidths, we selected the plasmon-vector q associated with the peak wavelengths from Fig. 6(e)–(g). Then, using the selected q, we plotted out the frequency linewidths in Fig. 7(a). It is observed that the frequency linewidths for graphene gap plasmons is small, with FWHM in the range of 1–2THz, compared to the larger linewidths for gold and silver, with FWHM in the range of 4–10THz. We also plotted out the gap plasmon power for gold and 0.4eV-graphene, in Fig. 7(b) and (c) respectively, for each plasmon vector q and photon energy. We see that the energy-spread for graphene gap plasmons is smaller, contributing to a higher peak-energy along the plasmon dispersion curve, while for gold the energy-spread is larger, contributing to a lower peak-energy. This finding demonstrates a potential direction to develop large tunability and high-Q nanoplasmonic devices based on the graphene platform for on-chip electrical-plasmon generation and manipulation.

The plasmon linewidth is partly governed by the relaxation time constant of the material, $\tau$, which is the mean free time of electronic collisions that leads to optical losses. In metals, $\tau$ is usually small, in the order of 0.01–0.1ps [22, 23], and thus optical losses are high. For graphene, $\tau$ is given by $\tau = \mu_e E_F / e v_F^2$ [26], where $\mu_e = 10^4 cm^2/V\text{-}s$ is the typical value of the carrier mobility [26, 27], and thus $\tau$ is evaluated to be in the range of 0.1–0.4ps for doped-graphene Fermi-levels from 0.1–0.4eV. Hence, the optical losses and plasmon linewidth could be made smaller with higher carrier mobilities and doping levels. Carrier mobilities are largely limited by charged-impurity scattering in the graphene sample [28]. While there are concerns that chemical doping of graphene would introduce impurities that degrade the carrier mobility, there are a few ways to overcome this. For example, molecular adsorption could act as compensators that neutralizes the effects of the impurities [29]. On the other hand, the Fermi-level of graphene could also be tuned by other methods, for example, electrostatic-gating and photo-induced doping, that preserve the carrier mobility [27, 30].

## V. Conclusions

In conclusion, we have studied the IETP excitation for graphene and metals. The generation efficiency of graphene plasmons is expected to be very much larger compared to metal plasmons due to the former's narrow plasmon linewidths and SPPs operating in the mid-infrared wavelength regime. The IETP excitation of graphene has the potential to be an efficient and low-powered plasmon source for graphene-based optoelectronic devices, which will show great promise in developing the field of on-chip electrical-plasmon generation and manipulation for nanophotonic integrated circuits.

## VI. Acknowledgments

This work is supported by the SUTD-MIT IDC grant (IDG21200106 and IDD21200103). H. S. Chu acknowledges the support of the National Research Foundation Singapore under its Competitive Research Programme (NRF-CRP 8-2011-07). L. K. Ang acknowledges the support of a USA AFOSR AOARD grant (14-4020).

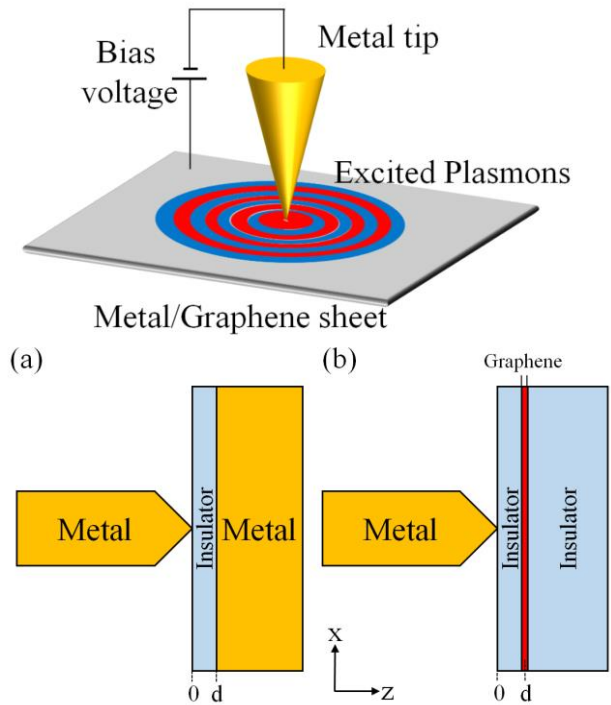

Fig. 1 Schematic of the plasmon excitation from a metal-insulator-metal/graphene tunnel junction. (a) Metal-insulator-metal tunnel junction. (b) Metal-insulator-graphene tunnel junction.

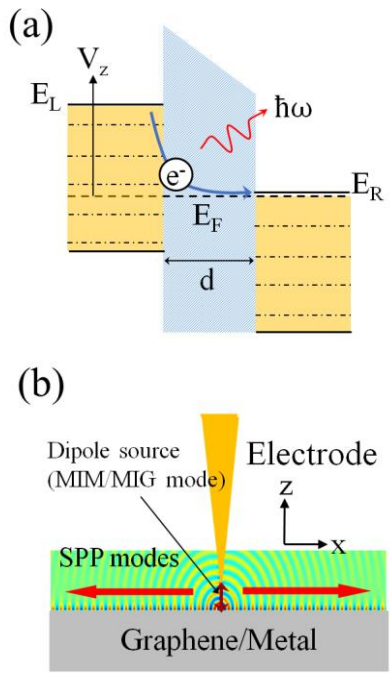

Fig. 2 (a) Mechanism of the inelastic electron tunnelling in generating a gap plasmon. On applying a bias-voltage $V_z$, the energy-level of the left electrode is shifted and electrons tunnel inelastically to the right electrode, exciting a gap plasmon with energy $\hbar\omega=eV_z$. (b) Model of the 2D FDTD simulation. The gap plasmon mode is modelled as a line dipole source with a normalized power of 1W at all wavelengths. The power is coupled out as SPP modes and propagate along the metal/graphene surface.

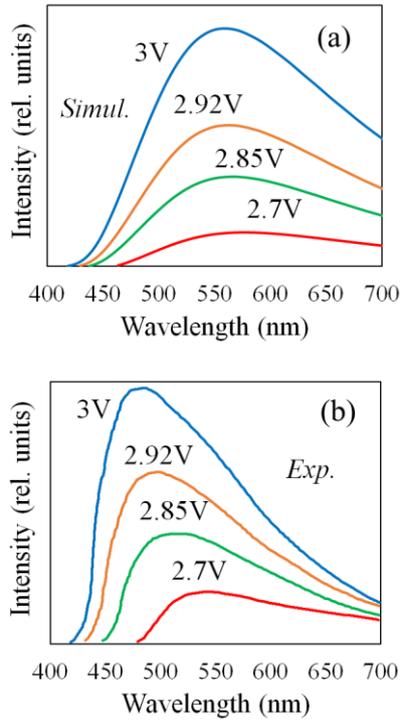

Fig. 3 Comparison of intensity spectrum of Al-Al$_2$O$_3$-Ag IETP system: (a) simulated and (b) experimental [4].

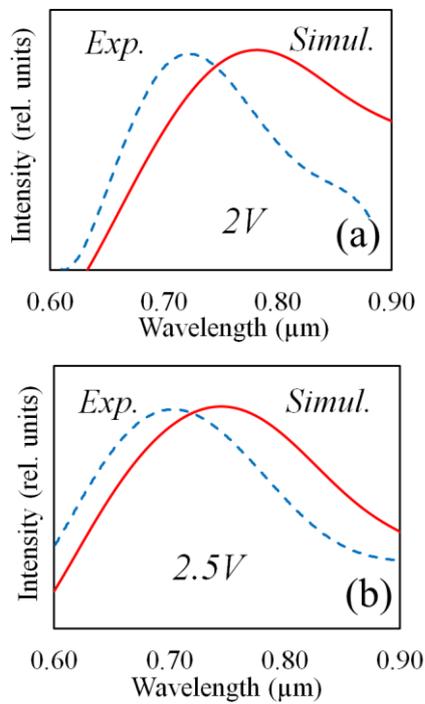

Fig. 4 Comparison of simulated (solid lines) and experimental (dashed lines) intensity spectrum of Au-air-Au IETP system under bias of (a) 2V [6] and (b) 2.5V [10].

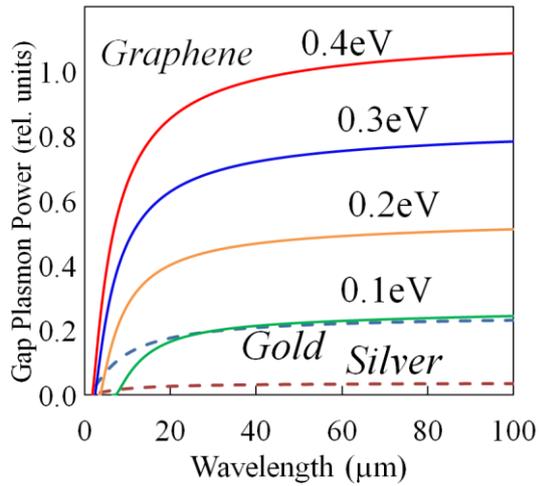

Fig. 5 IETP gap plasmons from tunnelling from an aluminium tip to metals (dashed lines), and to graphene of various Fermi-levels (solid lines).

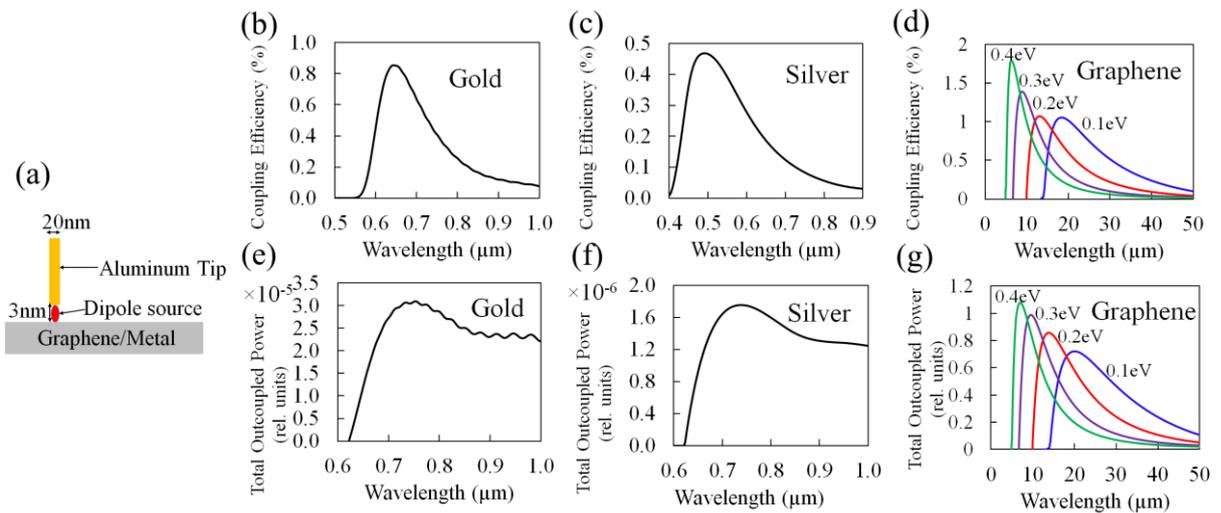

Fig. 6 (a) Schematic of the FDTD electromagnetic simulation structure. (b), (c) and (d) are coupling efficiencies of gap plasmons to gold, silver and graphene substrates respectively. (e), (f) and (g) are the total efficiencies in generating SPPs in gold, silver and graphene respectively.

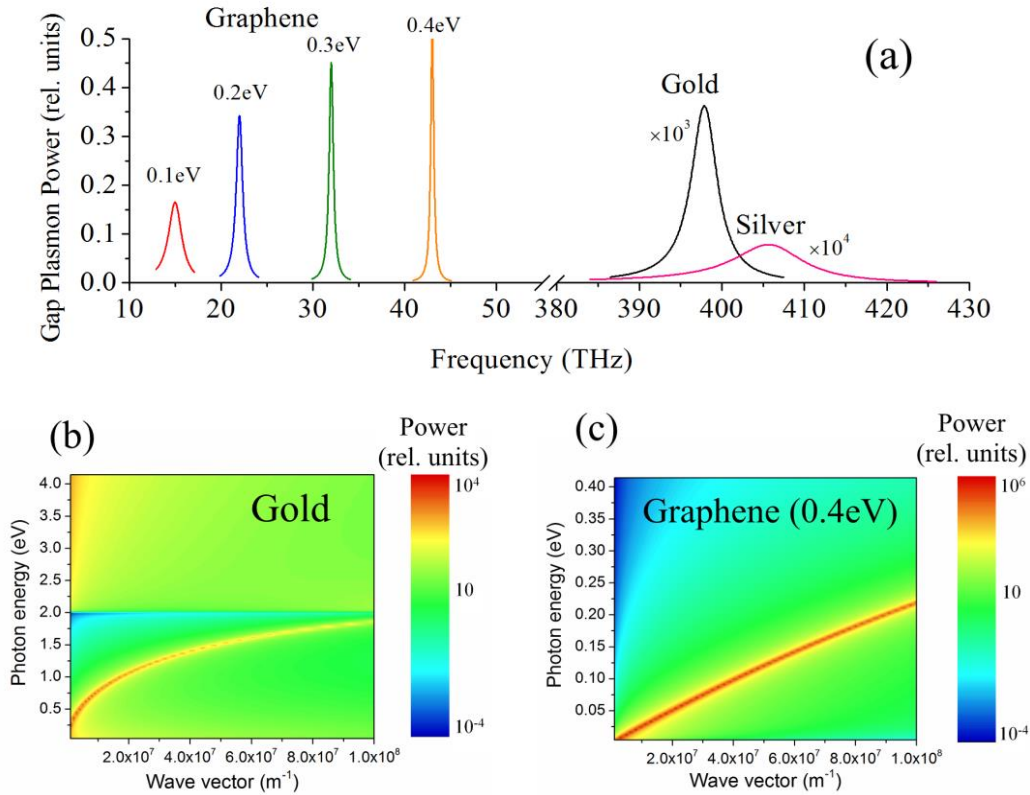

Fig. 7 (a) Frequency linewidths for the gap plasmons for graphene, gold and silver. Graphene gap plasmons have narrower linewidths compared to metal gap plasmons. (b) and (c) are gap plasmon power plots for each plasmon vector q and photon energy in eV, for gold and 0.4eV-graphene respectively. The gap plasmon power for graphene is more confined along the plasmon dispersion curve compared to gold. These results show that graphene exhibits much larger Q-factor and is tuneable compared to gold and silver materials.